%% file: 0-Abstract.tex
\documentclass[10pt,a4paper,twoside]{article}
\usepackage{JISE}
\usepackage{graphicx}
\usepackage{times}
\usepackage{mathptmx}
\usepackage{algorithm}
\usepackage{algorithmic}
\usepackage{url}
\usepackage{CJKutf8}

\title{Input Relation Prompting for Metamorphic Testing on Query-Based Systems
}
\author{Eng-Shen Tu and  Shin-Jie Lee\\
  {\small\em Department of Computer Science and Information Engineering} \\
  {\small\em National Cheng Kung University}\\
  {\small\em Tainan, Taiwan}\\
  {\small\em jamestu6301@gmail.com; jielee@mail.ncku.edu.tw}
}

\authorrunning{Eng-Shen Tu, Shin-Jie Lee}
\titlerunning{Input Relation Prompting for Metamorphic Testing }

\begin{document}
\editorfootnote{Received August 10, 2023; revised November 15, 2023; accepted December 20, 2023. \\
}
\maketitle
\begin{abstract}
  Testing query-based systems (QBSs) presents significant challenges due to the absence of ground truth for validation and the extensive time and effort required for manual testing. This paper addresses these challenges by proposing an approach that assists testers in identifying metamorphic relations (MRs) for metamorphic testing (MT) instead of solely and exhaustively relying on prerequisite domain knowledge. MT is an approach rising in popularity employed to alleviate the oracle problem by applying input transformation rules (MRs) to a program. The proposed approach helps the tester by prompting MRs that describe the relationships between inputs and outputs, enabling fault detection when the expected relationship is not met. Unlike traditional testing approaches, this approach does not rely on pre-defined test cases or concrete ground truth, making it suitable for the testing of real-world QBSs. Furthermore, the proposed approach can be combined with other testing methods such as combinatorial testing and fuzz testing, expanding the possibilities for QBS testing. A conducted case study of a real-world web application demonstrates the applicability and potential of the proposed approach. Overall, this research contributes to advancing the field of metamorphic testing and provides a valuable tool for QBS testers to enhance their testing efficiency.
\end{abstract}

\input{1-Introduction.tex}
\input{2-Methodology.tex}
\input{3-Experiments.tex}
\input{4-Discussion.tex}
\input{5-RelatedWork.tex}
\input{6-Conclusion.tex}

\section*{ACKNOWLEDGMENT}
This research is sponsored by the Ministry of Science and Technology under the grant NSTC 112-2221-E-006 -084 -MY2 in Taiwan.
\bibliographystyle{JISEbib}
\bibliography{references}

\begin{IEEEbiography}[{\includegraphics[width=3cm,height=4cm]{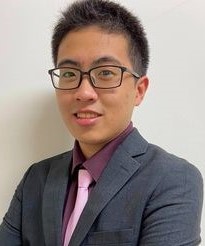}}]{Eng-Shen Tu \begin{CJK*}{UTF8}{bsmi}
(杜霙笙)
\end{CJK*}
} is a research assistant in the Software Engineering and Intelligent Test Automation Lab at National Cheng Kung University in Tainan, Taiwan. Currently, he is working towards his Bachelor's degree in Computer Science and Information Engineering at the same university. Though still early in his academic journey, James continues to explore and deepen his understanding of his chosen fields, always keen to contribute to new knowledge and innovative solutions in the domain of Information Science and Software Engineering.
\end{IEEEbiography}

\begin{IEEEbiography}[{\includegraphics[width=3cm,height=4cm]{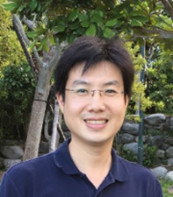}}]{Shin-Jie Lee \begin{CJK*}{UTF8}{bsmi}
(李信杰)
\end{CJK*}
} is an Associate Professor in Computer and Network Center at National Cheng Kung University (NCKU) in Taiwan and holds joint appointments from Department of Computer Science and Information Engineering at NCKU. His current research interests include software engineering and web test automation. He is the creator and team lead of SideeX project, serving as a basis for the most popular open source record-playback test automation tool in the world - Selenium IDE. He received his Ph.D. degree in Computer Science and Information Engineering from National Central University in Taiwan in 2007.
\end{IEEEbiography}

\end{document}

%% file: 1-Introduction.tex
\section{Introduction} \label{IntroductionSec}
Query-based systems (QBSs) are software applications that rely on query processing sub-systems to allow users to efficiently retrieve information from a database. These systems can range from simple search engines to complex business intelligence systems. The testing of these QBSs involves verifying that the system correctly processes user queries and returns the expected results. When testing QBSs, due to the immense volume of data and the complexity of QBSs, there are a number of challenges that arise when testing a given QBS. The two main challenges we would like to draw attention to are: 
\begin{itemize}
\item In QBSs, the correct output of the queries is either unknown or hard to compare with the observed output; therefore, it can be assumed that there is a lack of ground truth for most QBSs that require testing.\cite{8785455}\cite{weyuker1982testing} The ground truth mentioned here refers to a correct version of the software under test to refer to for comparison when testing for errors. The absence of ground truth would lead to the challenge of indistinguishable correct and incorrect behavior regarding the output of the QBS under test, which is known as the Oracle Problem\cite{6963470} in software testing. 
\item When given a QBS to test, manual testing and checking for errors or anomalies require an extensive amount of time and effort due to the vast volume of data. 
\end{itemize}

\begin{figure}[ht]
\centering
\includegraphics[width=\textwidth]{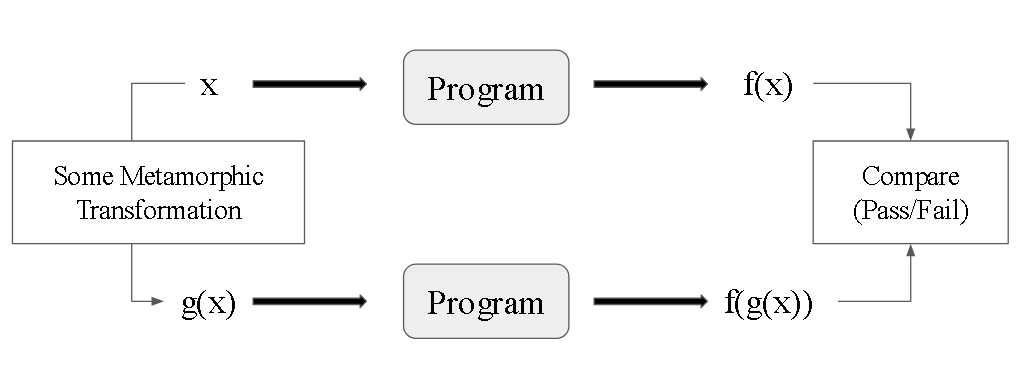}
\caption{Metamorphic Testing Flowchart.}
\label{fig1}
\end{figure}

Dealing with the first challenge, metamorphic testing (MT)\cite{chen2020metamorphic}, a software testing technique rising in popularity, is known to effectively alleviate the oracle problem\cite{6613484, 7422146} in multiple cases, including database systems\cite{7202957}. MT focuses on identifying faults in software by applying input transformation rules or metamorphic relations (MRs) to a program. It is based on the ideology that if two inputs are related by a known transformation, then the output of the program should also be related predictably, see Fig.\ref{fig1}. 

In a query-based system like a search engine, an example of metamorphic testing would be to alter a search query slightly and then compare the results. For instance, if the original query is "all universities in Taiwan", a metamorphic test might involve changing the query to "all universities in Asia" and then checking whether the search results of the first query are included in the second query since all universities in Taiwan are also universities in Asia. This tests the search algorithm's ability to recognize and correctly handle hierarchical or inclusive relationships between different sets of data. 

The goal of MT is to test the functionality of a software program without relying on a set of pre-defined test cases, which is well-suited for testing QBSs since not all QBSs have a wel. Instead, the tester defines a set of metamorphic relations that describe the relationships between different inputs and outputs. If the output does not match the expected relationship with the input, a fault is detected.

However, to effectively apply the use of MT\cite{chen2020metamorphic} in the testing of QBSs, one of the main challenges lies in MR identification\cite{10.1145/3143561}. The identification of MRs usually exhaustively relies on domain knowledge, which in most cases, might not be available without additional resources, especially when the tester is not well-versed with domain knowledge on the subject of the data. 

\begin{figure}[ht]
\centering
\includegraphics[width=\textwidth]{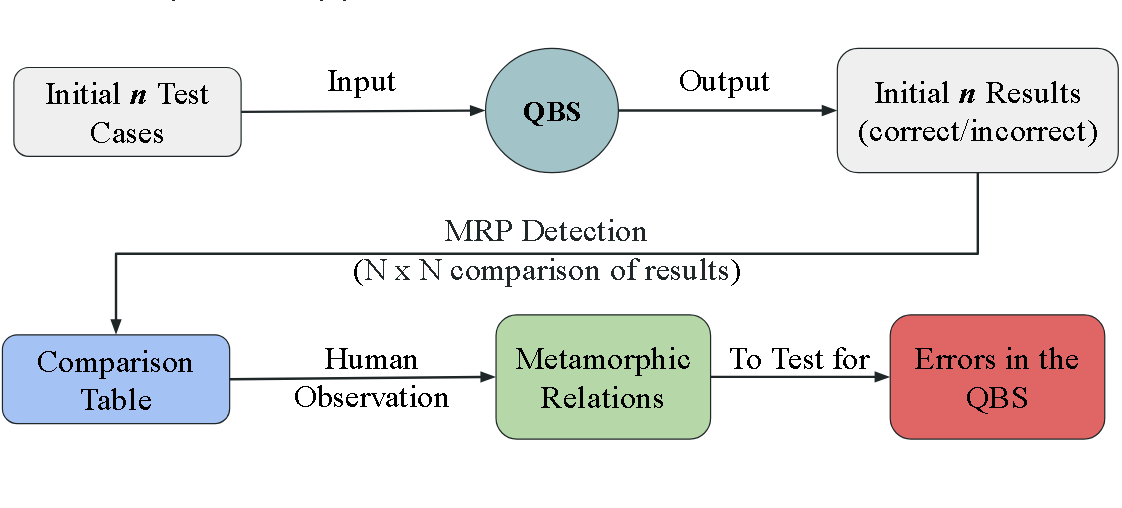}
\caption{The Proposed Approach.}
\label{overviewFig}
\end{figure}
Not all QBSs have well-defined or explicit specifications readily available when tested; therefore, this paper proposes an approach that supports real-world scenarios where an not all QBSs under test have ground truth, detailed specifications, or MRs available. The proposed approach will assist the process of identifying and defining MRs through prompting based on metamorphic relation patterns (MRPs) in the QBS and performing testing. The proposed approach relies less on prerequisite domain knowledge and does not require ground truth, making the approach more accessible and user-friendly to testers. Furthermore, the proposed approach can even be applied to and or combined with other testing methods, such as combinatorial testing (CT) and fuzz testing (FT). This increases the value of the proposed approach in terms of performing a more comprehensive test for errors on the given QBS. 

The organization of this paper is as follows. The MRPs will be outlined in Section \ref{MRPSec} Metamorphic Relation Patterns; details of the proposed approach will be described in Section \ref{ApproachSec} Methodology; and related work will be addressed in Section \ref{RelatedWorkSec} Related Work. The experiments of the proposed approach will be demonstrated via two case studies in Section \ref{ExperimentsSec} Experiments. The limitations and future work of the proposed approach will be discussed in Section \ref{DiscussionSec} Discussion and the conclusion will be drawn in Section \ref{ConclusionSec} Conclusion. 

\section{Metamorphic Relation Patterns} \label{MRPSec}
There are seven MRPs that will be frequently used and mentioned in this paper. These MRPs were introduced in detail by Segura et al.\cite{8785455} in 2019, and detailed examples for these MRPs can be found in both the original work\cite{8785455} and subsequent research\cite{9826124}. Here we provide a brief introduction for these seven MRPs, please note that the relational algebra notations used below are defined and specified in \cite{8785455}: 
\begin{itemize} \label{MRPs}
    \item Input Equivalence: 
    This pattern represents relations where source and follow-up test cases have equivalent inputs, resulting in the same output items in the same order.
    As mentioned in previous work \cite{8785455}, let $q_{1}= \mathcal{T}(\sigma(R,\ c_{1}))$ and $q_{2}=\mathcal{T}(\sigma(R,\ c_{2}),\ \langle o_{2}\rangle)$ be two queries defined under the same relation $R$. Input equivalence states that the outputs of $q_{1}$ and $q_{2}$ should be the same when conditions are equivalent, $c_{1} \equiv c_{2}$ , and ordering sequence is equivalent, $o_{1} \equiv o_{2}$, i.e.,
    \begin{align*} (c_{1} & \equiv c_{2}\wedge o_{1}\equiv o_{2})\Leftrightarrow\\ & (\mathcal{T}(\sigma(R,\ c_{1}),\ \langle o_{1}\rangle)= \mathcal{T}(\sigma(R,\ c_{2}),\ \langle o_{2}\rangle)) \end{align*}
    \item Shuffling: This pattern represents relations where the source and follow-up outputs contain the same items regardless of the ordering criteria specified as input.
    As mentioned in previous work \cite{8785455}, let $q= \sigma(R, c)$ be a query such that its result contains several attributes ${\{a_{i}\}}_{t=1, n}$ that can be used as ordering criteria with the $\mathcal{T}$ operator. Shuffling states that the result of the query $q$ ordered by a given attribute $a_{i}$ should contain the same elements as the same query ordered by any other attribute $a_{j}$, i.e.,
    \begin{align*} \forall i,j: & 1\ldots n\vert i\not= j\bullet\\ & items \mathcal{T} (q, \{a_{i}\rangle)= items \mathcal{T} (q, \langle a_{j}\rangle) \end{align*}
    \item Conjunctive Conditions: 
    This pattern involves refining a query with additional conjunctive conditions, where the results of each test case should be included in the previous ones.
    As mentioned in previous work \cite{8785455}, let $c$ be a complex condition formed by the conjunction of simpler conditions, i.e., $c=\mathop{\wedge}_{t=1\ldots n}c_{i}$. Conjunctive conditions states that the result of a query after one condition is added conjunctively to its selection condition should be a subset of the original query, i.e., 
    \begin{equation*} \forall i = 2\cdot \cdot n\bullet \sigma(R,\bigwedge_{k= 1\cdot \cdot i}\ldots c_{k})\subseteq \sigma(R,\mathop{\wedge}_{k= 1\ldots i-1} c_{k}) \end{equation*}
    \item Disjunctive Conditions: 
    This pattern involves expanding a query with disjunctive conditions, where the results of each test case should be a subset of the following ones.
    As mentioned in previous work \cite{8785455}, let $c$ be a complex condition formed by the disjunction of simpler conditions, i.e., $c=V_{i=1\ldots n}c_{i}$. Disjunctive conditions states that the result of a query when one condition is added disjunctively to its selection condition should contain the result of the original query, i.e., 
    \begin{equation*} \forall i=2\ldots n\bullet\sigma(R,\ \bigvee_{k=1\ldots i-1} c_{k}) \subseteq\sigma(R, \bigvee_{k=1\ldots i}c_{k}) \end{equation*}
    \item Disjoint Partitions: 
    This pattern represents relations where the outputs of follow-up test cases should have no items in common because the queried relation can be partitioned based on input attribute values.
    As mentioned in previous work \cite{8785455}, let $q=\sigma(R,c)$ be a query such that its result contains at least one attribute $a_{p}$ whose domain is a discrete set of values, e.g., $\{v_{1}, v_{2}, \ldots v_{n}\}$. Disjoint partitions states that, when a conjunctive condition of the form $a_{p}=v_{i}$ appears,  the result of two queries should be disjoint if the values with which $a_{p}$ is compared are different, i.e., 
    \begin{align*} \forall i,j &: 1\ldots n\vert i\neq j\bullet\\ & \sigma(R,\ c\wedge a_{p}=v_{i})\cap\sigma(R,\ c\wedge a_{p}=v_{j})=\oslash \end{align*}
    \item Complete Partitions: 
    This pattern represents relations where the union of follow-up outputs should contain the same items as the source output because the queried relation can be partitioned based on input attribute values.
    As mentioned in previous work \cite{8785455}, let $q=\sigma(R,c)$ be a query such that its result contains at least one attribute $a_{p}$ whose domain is a discrete set of values, e.g., $\{v_{1}, v_{2}, \ldots v_{n}\}$. Complete partitions states that the result of $q$ should be equal to the union of the results of the $n$ queries formed by a conjunctive condition of the form $a_{p}=v_{i}$ to the condition in $q$ for each possible value of $a_{p}$, i.e., 
    \begin{equation*} \sigma(R,\ c)=\bigcup_{i=1.n}.\sigma(R,\ c\wedge a_{p}=v_{i}) \end{equation*}
    \item Partition Difference: 
    This pattern represents relations where the outputs of follow-up test cases are pairwise disjoint, and their union contains the same items as the source output because the queried relation can be partitioned based on input attribute values.
    As mentioned in previous work \cite{8785455}, let $q=\sigma(R,c)$ be a query such that its result contains at least one attribute $a_{p}$ whose domain is a discrete set of values, e.g., $\{v_{1}, v_{2}, \ldots v_{n}\}$. Partition difference states that the difference between $q$ and the union of the results of the $k$ queries formed by adding a conjunctive condition of the form $a_{p}=v_{i}$ to the condition in $q$ for $k$ different possible values of $a_{p}$, should be equal to the result of the union of the results of the $n- k$ queries formed by adding a conjunctive condition of the form $a_{p}=v_{i}$ to the condition in $q$ for the $n- k$ different possible values of $a_{p}$, i.e., 
    \begin{align*} \forall k:1\ldots (n- & 1)\bullet\\ \sigma(R,\ c)- & \bigcup_{i=1.k}.\sigma(R,\ c\wedge a_{p}=v_{i})=\\ & \bigcup_{j=k+1}\ldots \sigma(Rn,\ c\wedge a_{p}=v_{j}) \end{align*}
\end{itemize}

%% file: 2-Methodology.tex
\section{Methodology} \label{ApproachSec}
\subsection{Overview}
Our proposed approach is focused on helping the tester discover and identify input relations by prompting them with metamorphic relation patterns (MRPs)\cite{8785455} detected within the QBS to perform metamorphic testing and check for errors or anomalies in a given QBS. 

Based on real-world scenarios for QBS testing, with a given QBS, there is usually no ground truth, hence the need for testing. In this view, when faced with the task of testing the given QBS through querying, testers encounter the Oracle problem\cite{6963470} and do not have correct results for validation. This implies that with a given set of initial test cases, the queried results are a mix of correct or incorrect results. 

The proposed approach theorizes that if at least some portion of the queries yields accurate results, it should be possible for the tester to discover input relations when prompted with MRPs\cite{8785455} that can be detected between the results of the given test cases. This works even if only some of the results are correct since MRs should hold true consistently within the QBS, so if there lies a logical explanation, it is indeed possible to identify these MRs when prompted with partially correct results. 

For the proposed approach, the MRPs for QBSs\cite{8785455} previously mentioned in Sec. \ref{MRPs} prompt the tester to identify MRs within the QBS under test. 

\begin{figure}[ht]
\centering
\includegraphics[width=\textwidth]{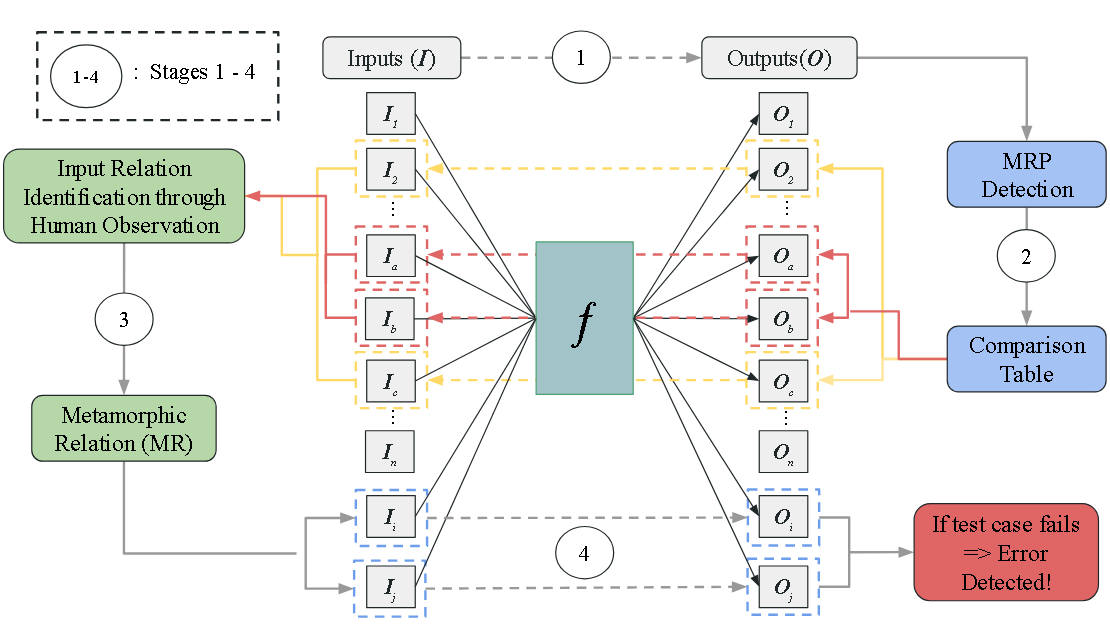}
\caption{Operational Concept of the Proposed Approach.}
\label{approachFig}
\end{figure}

\subsection{The Proposed Approach}
Given a QBS to test and an $n$ number of test cases, later described in Section 2.2.1, the QBS is queried to yield $n$ initial results, keeping in mind that the initial $N$ results could contain both correct and incorrect results. Next, MRP detection is performed for the generated results which then creates a comparison table, which specifies MRP relations detected between the $n$ results, the process of comparison table generation is shown in Section \ref{Stage2sec}. With the comparison table, the tester can then observe and identify input relations for MT and potentially find errors and or anomalies in the QBS. The overview of the proposed approach can also be seen in Fig. \ref{overviewFig}. 

In the following subsections, a more detailed description as well as a running example of the proposed approach will be given in accordance with Fig. \ref{approachFig}, which contains a detailed visualization of the operational concept of the proposed approach when given a QBS to test and an initial set of test cases. 

\subsubsection{\textbf{Obtaining Initial Test Case Results}} \label{Stage1sec}
In Fig. \ref{approachFig}, $f$ denotes the function of querying the QBS under test, $I$ denotes the initial test cases input, $O$ denotes the queried output of $I$, $n$ denotes the total number of initial test cases, and the following equation, Eq. (\ref{getResults}), shows the input and output relation function of querying the QBS under test:  
\begin{equation} 
    \label{getResults}
	f(I_i) = O_i
\end{equation}
where $I_i$ denotes an input test case and $O_i$ denotes the corresponding output test case. Referencing Stage 1 in Fig. \ref{approachFig}, given the initial test cases $I$ as input, the QBS under test is queried with the initial test cases, and the results $O$ are saved for comparison in the next step. 

For the initial test cases, the proposed approach shows high flexibility and compatibility, which implies that the proposed approach can work well with numerous types of test cases generated through other testing methods. However, it is still worth noting that the better designed the initial test cases are, the easier it will be for the tester to identify input relations. In this context, test cases targeted toward singular or pairwise parameters typically yield more obvious prompts to the tester compared to test cases that cover a complex combination of multiple parameters. 

For example, given the following hypothetical QBS and initial test cases, obtaining the initial test case results would yield:
\begin{itemize}
    \item QBS Under Test: An Animal Image Search Engine
    \item Initial Test Case Queries and their Outputs: 
    \begin{itemize}
        \item Cats: All cat images in the system. 
        \item Funny Birds: All funny bird images in the system. 
        \item Dogs: All dog images in the system. 
        \item Pandas: All panda images in the system. 
        \item Funny Dogs: All funny dog images in the system. 
        \item Funny Cats: All funny cat images in the system. 
        \item Birds: All bird images in the system. 
        \item Cute Dogs: All cute dog images in the system. 
    \end{itemize}
\end{itemize}
We will be following this running example for each step. 
\begin{figure}[ht]
\centering
\includegraphics[width=\textwidth]{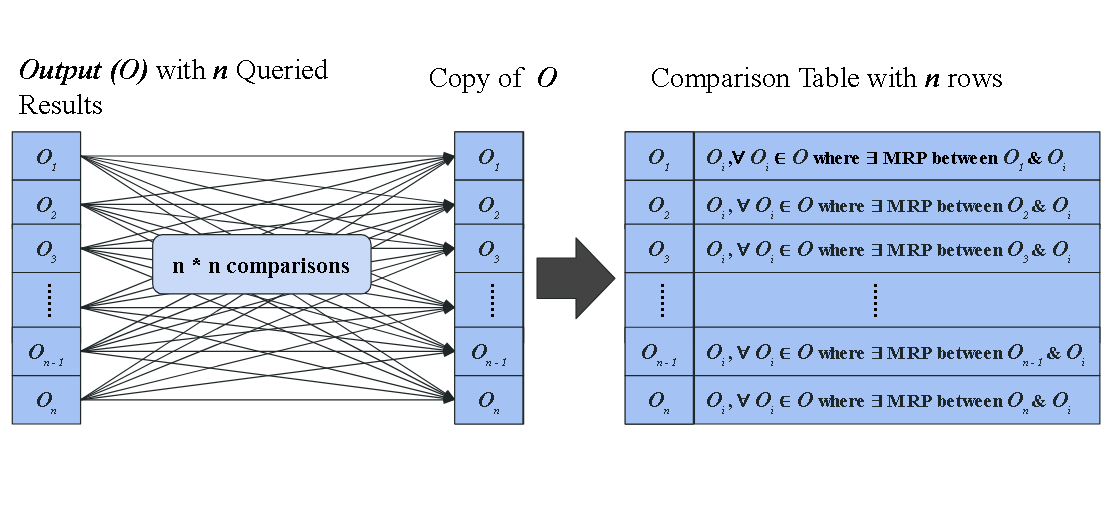}
  \caption{Visualization of Comparison Table Generation.}
  \label{CTDFig}
\end{figure}

\subsubsection{\textbf{MRP Detection}} \label{Stage2sec}
With $O$ from Sec. \ref{Stage1sec}, the comparison table is then generated by performing MRP Detection. Referencing Stage 2 in Fig. \ref{approachFig}, MRP Detection is conducted by making comparisons between all pairwise combinations of $O$, and the results of these comparisons will be held in the generated comparison table, see Fig. \ref{CTDFig} for visualization of the process. 

As for the details of comparison table generation, the pseudocode for the process is provided, see Algorithm \ref{Algo1}, in which the comparison table is returned for the next stage in the proposed approach. In the implementation of the proposed approach, the comparison table is created in the form of a data frame and saved to an Excel file for observation. 

\begin{algorithm}
\caption{Comparison Table Generation}
\label{Algo1}
\begin{algorithmic}
\STATE $results \gets $ table of results of initial test cases
\STATE $comparisonTable \gets \emptyset$ 
\FORALL{$queryResult_a \in results$} 
\FORALL{$queryResult_b \in results$} 
\IF{MRP detected in $queryResult_a \And queryResult_b$}
\STATE $a \gets $ index of $queryResult_a $ in results
\STATE $b \gets $ index of $queryResult_b $ in results
\STATE add $comparisonTable[a]$ to $results[b]$ 
\ENDIF
\ENDFOR
\ENDFOR
\RETURN $comparisonTable$
\end{algorithmic}
\end{algorithm}

Following the running example, while testing for the MRP conjunctive condition in Section.\ref{MRPs}, we should be able to build the comparison table through comparing the results. From the generated comparison table Table.\ref{table:cte}, we can see that the output of the test cases Funny Birds, Funny Dogs, Funny Cats, and Cute Dogs are subsets of the queries Birds, Dogs, Cats, and Dogs respectively. Please note that this process will be done for all MRPs in Section.\ref{MRPs}. 

\begin{table}
    \centering
    \begin{tabular}{|c|c|} \hline 
         Queried Results ($O$)& Conjunctive Conditions\\ \hline 
         Cats& Funny Cats\\ \hline 
         Funny Birds& None\\ \hline 
         Dogs& Funny Dogs, Cute Dogs\\ \hline 
         Pandas& None\\ \hline 
         Funny Dogs& None\\ \hline 
         Funny Cats& None\\ \hline 
         Birds& Funny Birds\\ \hline 
         Cute Dogs& None\\ \hline
    \end{tabular}
    \caption{Comparison table of running example. }
    \label{table:cte}
\end{table}
\subsubsection{\textbf{Human Observation}} \label{Stage3sec}
With the generated comparison table, the tester can observe the MRP relations between the outputs in $O$, map them back to their respective inputs in $I$, and identify potential input relations, which serve as the key to performing MT. Referencing Stage 3 in Fig. \ref{approachFig}, prompted by the comparison table generated from Sec. \ref{Stage2sec}, if there exists an intuitive logical explanation behind the input relation identified, then it is likely that an MR can be defined for the QBS. 

In this stage, even though the tester does not need to rely on prerequisite domain knowledge extensively to identify these input relations of the QBS, the tester could be required to make assumptions about what the input relations could be. However, with the proposed approach, we claim that being prompted by MRP detection and the generated comparison table makes this task significantly easier. Also, since the comparison table is stored in the form of a data frame so it is also possible for the tester to manipulate the generated comparison table to aid the process of input relation identification. 

In the running example, from human observation, we can intuitively infer that this could be because all funny birds are birds, all funny dogs are dogs... and so on. Therefore, through this process, we are able to assume and identify an input relation, that is, the results of the query with keyword $funny$ followed by any given animal $A$ should be a subset of the results of the query $A$. Next, we will then be able to perform MT with this input relation on all animals in the QBS. 

\subsubsection{\textbf{Testing for Errors}}
Once MRs have been identified and defined, these MRs can then be used to perform MT\cite{chen2020metamorphic} on the QBS. Referencing Stage 4 in Fig. \ref{approachFig}, for each MR, for all inputs $I_i$ and $I_j$ that fit the input relation, their corresponding outputs $O_i$ and $O_j$ are checked to see if they fulfill the output relations defined by the MR. If they do not fulfill the defined relation, then the test case fails and an error will have been detected within the QBS under test. 

Furthermore, it is worth noting that inputs $I_i$ and $I_j$ do not necessarily have to be within the initial test cases, they could be newly generated test cases that fit the relation specified by the identified MR to perform MT\cite{chen2020metamorphic} on the QBS under test. 

For the running example, if test cases arise where the input relation does not hold, e.g. the query $funny$ $pandas$ includes an image that is not a panda, which leads to the result of the query not being a subset of the query $pandas$, then that means we have found a potential error in the system. 

%% file: 3-Experiments.tex
\section{Experiments} \label{ExperimentsSec}
\begin{figure}[ht]
\centering
\includegraphics[width=\textwidth]{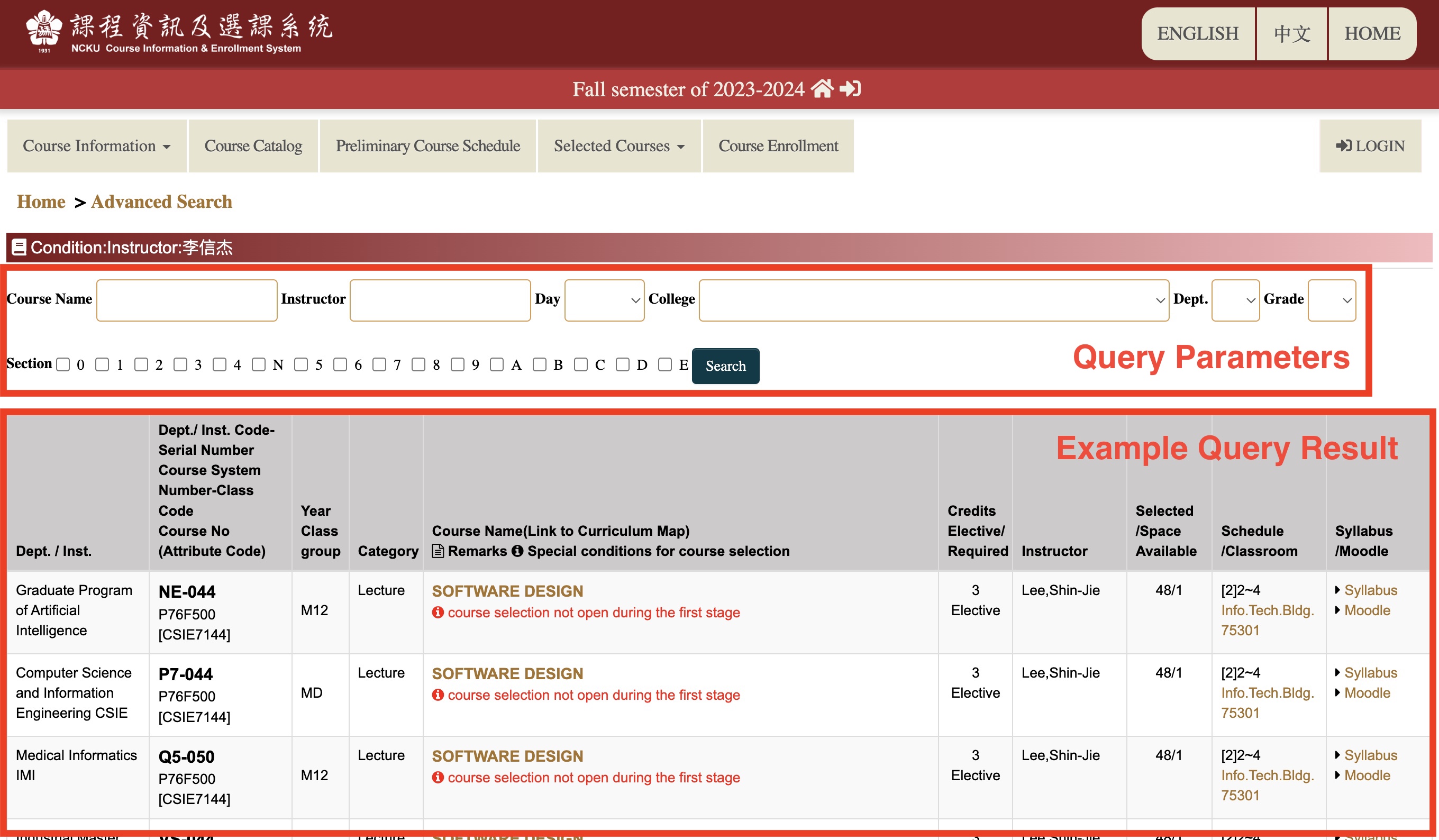}
\caption{The National Cheng Kung University Course Search Webpage\cite{nckucourseweb}.}
\label{nckuweb}
\end{figure}

\subsection{Experiments Overview}
For the proposed approach, the experiments were conducted in the form of a case study, applying the proposed approach combined with widely known test case generation approaches to a given real-world QBS. 

\textbf{\emph{The QBS Under Test: }}The experiments were performed on the National Cheng Kung University Advanced Course Search Website\cite{nckucourseweb}, a real-world QBS with no ground truth of query results available. This website belongs to National Cheng Kung University (NCKU) and is frequently used by students of NCKU to perform search queries for the courses of the current semester. In Fig. \ref{nckuweb}, we can see the structure of the website as well as the query parameters and an example query result. For the following case study, the values of the column: Dept/Inst Code Serial Number Course System Number represent the result of each query. 

\textbf{\emph{Fuzz Testing Case Study: }}In this case study, the proposed approach is applied to FT-generated test cases. FT is a widely known software testing technique that involves providing unexpected and random inputs to a program to identify vulnerabilities or bugs. It helps uncover potential issues by stressing the system with invalid or unexpected data, improving overall software security and reliability, as seen in practice in \cite{10.1145/3243734.3243804}\cite{8330260}. For this case study, the FT test cases are targeted toward the "Instructor" query parameter. 

\textbf{\emph{Combinatorial Testing Case Study: }}In this case study, the proposed approach is applied to CT-generated test cases. CT is a commonly used method in software testing with the overarching aim to minimize the number of test cases. By selecting a subset of possible combinations rather than testing every single combination, CT reduces the number of test cases required while still achieving high test coverage. CT can help identify interaction-related faults and improve testing efficiency. Applications of CT can be seen in \cite{10.1145/1883612.1883618}. For this case study, the combinatorial test cases are targeted toward the query parameters "Instructor" and "Day". 

\begin{figure}[ht]
\centering
\includegraphics[width=\textwidth]{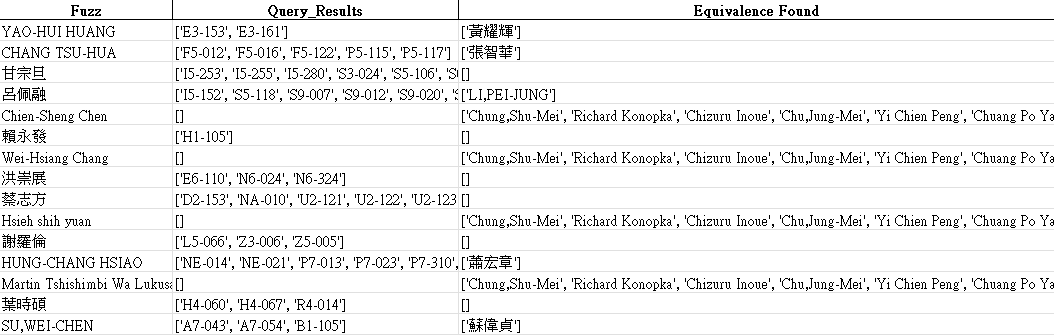}
\caption{Fuzz Testing Case Study: Comparison Table for Input Equivalence.}
\label{cs1_comp}
\end{figure}

\subsection{Fuzz Testing Case Study}
\subsubsection{Test Case Generation and Initial Querying}

In this experiment, FT was applied to test case generation for the "Instructor" search bar on the QBS webpage. During test case generation, a web crawler was scripted to get all the unique values listed in the instructor column of the QBS, which should include all names of teachers in NCKU. 

Once the crawled data is saved in the form of test cases, the QBS is then queried with these test cases, and the results are then saved for the next stage of the approach, MRP detection and comparison table generation. 

\subsubsection{MRP Detection and Comparison Table Generation}

In this case study, despite being able to generate comparison tables for all MRPs, only the comparison table for the MRP Input Equivalence is demonstrated, as it showed the most prominent results, see Fig. \ref{cs1_comp}, the rest are excluded for concision. Following Eq. (\ref{getResults}), the MRP input equivalence condition equation is defined as:
\begin{equation} \label{mrp_IE}
	f(I_i) \equiv f(I_j)
\end{equation}
where $I_i$ and $I_j$ in Eq.( \ref{mrp_IE}) are said to be input equivalent. 

In Fig. \ref{cs1_comp}, the Fuzz column contains all the names of teachers in National Cheng Kung University, the Query Results column holds the queried results, and the Equivalence Found column holds a list of inputs within the initial test case that yield an equivalent result to the Fuzz input. 

\begin{figure}[ht]
\centering
\includegraphics[width=\textwidth]{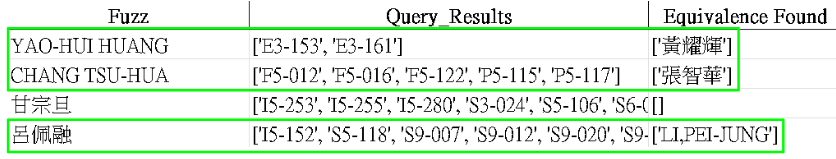}
\caption{Fuzz Testing Case Study: Observation of Teacher Name Input Relation.}
\label{cs1_tnr}
\end{figure}

\subsubsection{Input Relation Identification} \label{subsubSec_IRI}
Upon observation of the input equivalence comparison table, it is not hard to find an input relation between some of the teacher names. In Fig. \ref{cs1_tnr}, it can be observed that given the instances where there is only one case of equivalence, for each instructor, there is an equivalence between his or her English and Chinese name, which is logical since querying an Instructor's English name should return the same results as querying their Chinese name. This prompts the identification of an input equivalence MR equation for the Instructor parameter, which, following Eq. (\ref{getResults}) and Eq. (\ref{mrp_IE}), can be defined as: 
\begin{equation} \label{cs1_mr}
	f(TeacherEnglishName) \equiv f(TeacherChineseName).
\end{equation}

This MR can then be used to test the QBS\cite{nckucourseweb} under test in the next stage.

\begin{figure}[ht]
\centering
\includegraphics[width=\textwidth]{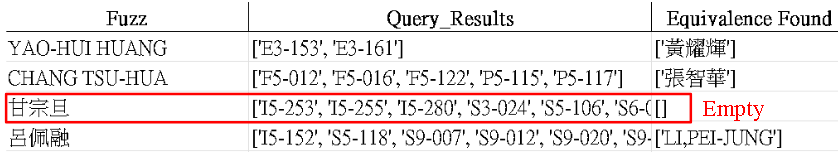}
\caption{Fuzz Testing Case Study: Errors found in the QBS under test\cite{nckucourseweb}.}
\label{cs1_error}
\end{figure}

\subsubsection{Check for Errors in QBS}
Now that the Eq. (\ref{cs1_mr}) has been induced, as an MR of the QBS under test\cite{nckucourseweb}, MT\cite{chen2020metamorphic} can now be performed to check for errors. In this case, since all the teachers' names were scraped and added during initial test case generation, no new test cases need to be generated to perform testing, and the results of the initial test cases can be conveniently used to validate the QBS under test. To check for errors, the results are validated by checking if there is an English Name counterpart to the Chinese Name and vice versa in the Equivalence Found column.  

\subsubsection{\textbf{Results and Remarks}}
After testing, errors were indeed found in the QBS under test, e.g. in Fig. \ref{cs1_error}, it is obvious there is no English Name counterpart to the Chinese Name in the Fuzz Column. Similar errors can also be found in Fig. \ref{cs1_comp}, and after complete validation, approximately half of the test cases exhibit this error. This finding can subsequently be reported to the owner of the website. 

In addition, it is also worth noting that even though the error in Fig. \ref{cs1_error} can be visually seen in Fig. \ref{cs1_tnr}, it does not substantively affect the tester's judgment when prompted by the comparison table to identify the input relation Eq. (\ref{cs1_mr}) mentioned in Sec. \ref{subsubSec_IRI}. 

Through this experiment, it is shown that the proposed approach can indeed be used to discover metamorphic relations and detect errors in QBSs even when the initial test cases do not yield completely correct results. Furthermore, this experiment also shows the compatibility and successful combination of the proposed approach and FT. 

\begin{figure}[ht]
\centering
\includegraphics[width=\textwidth]{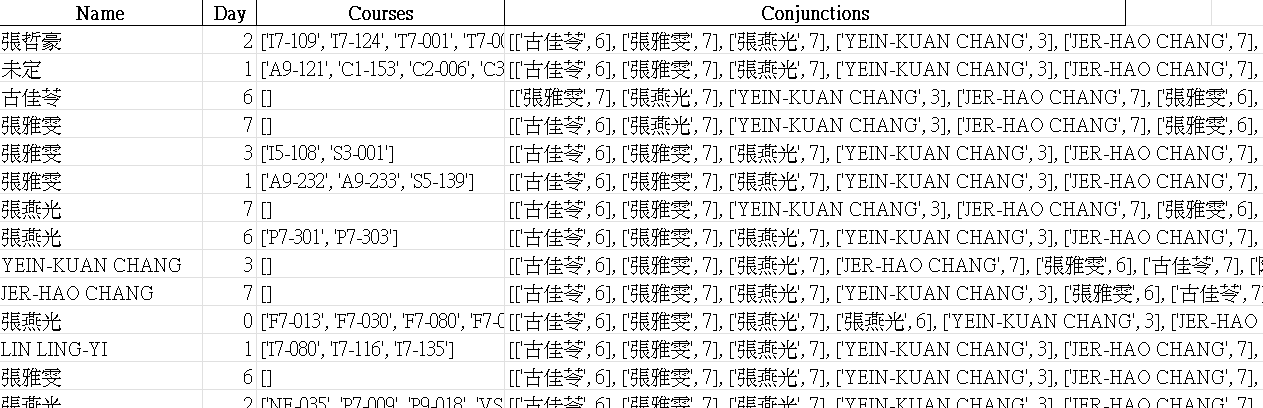}
\caption{Combinatorial Testing Case Study: Comparison Table for Conjunctions.}
\label{cs2_comp}
\end{figure}

\begin{figure}[ht]
\centering
\includegraphics[width=\textwidth]{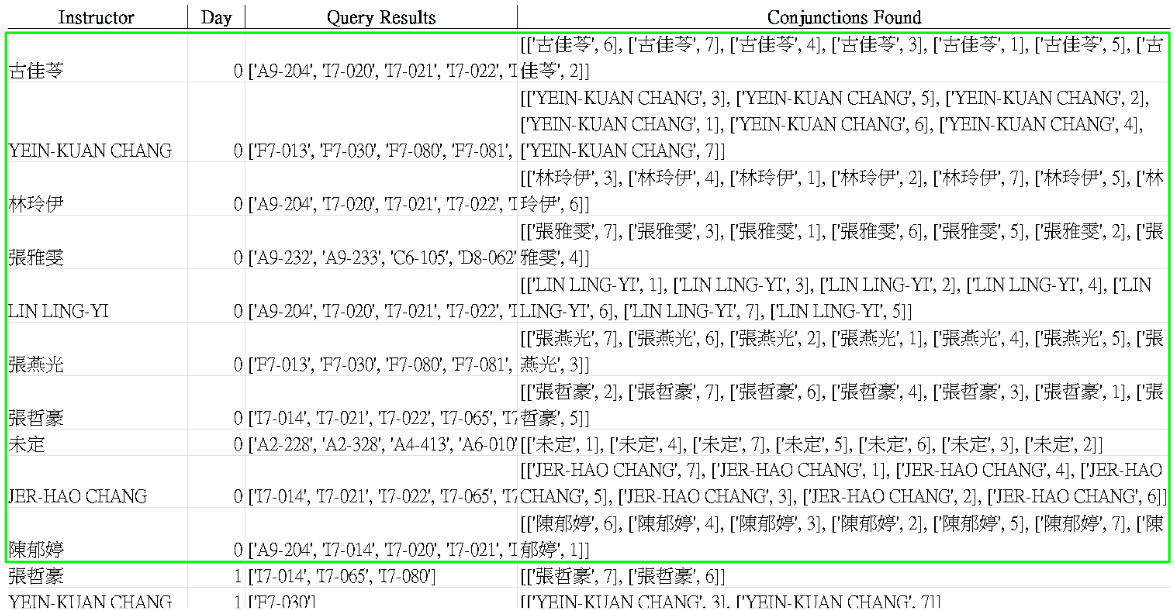}
\caption{Combinatorial Testing Case Study: Filtered and Sorted Comparison Table.}
\label{cs2_compF}
\end{figure}

\subsection{Combinatorial Testing Case Study} \label{CS2}
\subsubsection{Test Case Generation and Initial Querying} \label{cs2_tcg}
In the second experiment, the proposed approach is applied to CT. For the combinatorial test cases, the parameters targeted are the "Instructor" and "Day" parameters, see Fig. \ref{nckuweb}, in the QBS under test \cite{nckucourseweb}. For combinatorial test case generation, PICT\cite{pict} was used to generate pairwise test cases for the query parameters, specified as follows:
\begin{itemize}
    \item Instructor: A total of 10 instructors, each with more than $25$ courses in the QBS, were chosen. 
    \item Day: A value from $0-7$, with $1-7$ denoting the 7 days of the week and 0 defined as no selection of the day, which yields results of $1-7$ combined. 
\end{itemize}
With $10$ instructors and $8$ days, $80$ test cases were generated with PICT\cite{pict}, and the QBS under test\cite{nckucourseweb} was queried with all $80$ pairwise combinations. 

\subsubsection{Comparison Table Generation}
For the demonstration of the results of this experiment, only the comparison table of the MRP Conjunctive Conditions, see Fig. \ref{cs2_comp}, is shown. However, while the other MRP comparison tables are excluded for concision, it is worth noting that the MRPs Input Equivalence, Disjoint Partitions, Complete Partitions, and Partition Difference all yield well-performing MRs that can be identified with this set of test cases. In this case study, following Eq. (\ref{getResults}), the conjunctive condition equation is defined as: 
\begin{equation} \label{mrp_CC}
	j \in f(I_i) \text{, } \forall \text{ } j \in f(I_j)
\end{equation}
where $I_j$ is a subset of $I_i$. 

In Fig. \ref{cs2_compF}, the Instructor and Day columns contain the query parameters mentioned in Sec. \ref{cs2_tcg}, the Query Results column contains the course results of the QBS under test\cite{nckucourseweb}, and the Conjunctions Found column specify the subsets of the query found within the test cases, which are listed in the form [Instructor Name, Day]. 

\subsubsection{Input Relation Identification} 
The original Conjunctions comparison table, see Fig. \ref{cs2_comp}, is visually messy and complex, making it difficult to prompt the identification of any input relations. However, once the Conjunctions column is filtered by name and the data frame is sorted by Day, the resulting filtered and sorted comparison table can be seen in Fig. \ref{cs2_compF}, which is significantly more readable. 

Upon observation of the filtered and sorted comparison table in Fig. \ref{cs2_compF}, it can be observed that for each instructor, [Instructor Name, Day=[1-7]] are subsets of the test case [Instructor Name, Day=0], which can be used to prompt the identification of an MR for Conjunctive Conditions. The MR equation, following Eq. (\ref{getResults}) can be defined as: 
\begin{equation} \label{cs2_mr1}
	f(Name, Day=[1-7]) \in f(Name, Day=0).
\end{equation}

Upon further observation, since Day=0 was defined as the union of Days $1-7$, this observation can also be used to define an MR for Complete Partitions. This second MR equation can be defined as: 
\begin{equation} \label{cs2_mr2}
	f(Name, Day=0) \equiv \cup (f(Name, Day=[1-7])).
\end{equation}

These MRs can then be used to test the QBS\cite{nckucourseweb} under test in the next stage.

\subsubsection{Check for Errors in QBS}
With the induced MR equations, Eq. (\ref{cs2_mr1}) and Eq. (\ref{cs2_mr2}), as MRs of the given QBS\cite{nckucourseweb}, MT\cite{chen2020metamorphic} is performed to check for errors. In this case study, all test cases for both MRs passed, and no errors were detected. 

\subsubsection{\textbf{Results and Remarks}}
Even though no errors were found with these identified MRs in the QBS under test\cite{nckucourseweb}, it can be inferred from this result that Eq. (\ref{cs2_mr1}) and Eq. (\ref{cs2_mr2}) are MRs that hold consistently within the QBS under test\cite{nckucourseweb}. 

Furthermore, This experiment shows the compatibility and successful combination of the proposed approach and CT and even goes on to show that the proposed approach can help alleviate the oracle problem that CT might be challenged with when testing QBSs. 

%% file: 4-Discussion.tex
\section{Discussion} \label{DiscussionSec}
\subsection{Limitations}
Though versatile and arguably applicable to real-world scenarios, the proposed approach does have limitations that affect its performance. 
\begin{itemize}
    \item The proposed approach relies on the assumption that at least some portion of the queries of the initial test cases are correct. That is, if the portion of correct queries is too low, the tester may not be able to infer the input relations when prompted by the generated comparison table. 
    \item Following the previous limitation, if there are $0$ correct queries in the initial test case, then the proposed approach will fail to prompt correct MRs to the tester, nor will it be able to detect whether such errors that exist within the initial test cases. 
\end{itemize}

\subsection{Future Work}
This research unlocks intriguing research opportunities and challenges waiting to be addressed, including: 
\begin{itemize}
    \item Automation: Future work in this regard could be focused on how to automate the process that requires human observation and intervention to provide a fully automated solution. 
    \item Combining with other testing methods: In this paper, the case studies demonstrate the proposed approach combined with FT and CT; however, there could be other combinations that may yield valuable insight. 
    \item Application in other domains: This paper specifically focuses on QBSs; however, the proposed approach could have the potential to benefit other software systems beyond QBSs where MRPs can be identified. 
\end{itemize}

%% file: 5-RelatedWork.tex
\section{Related Work} \label{RelatedWorkSec}
In this section, we will address some related work as well as how this paper contributes to new findings. The related work presented in this section mainly focuses on metamorphic relation generation and metamorphic testing. 

\subsection{Metamorphic Relation Patterns for QBSs}
In 2019, Segura et al.\cite{8785455} presented a catalog of metamorphic relation patterns for Query-Based Systems\cite{8785455} from their observation that the MRs used to test different types of QBSs are very similar, regardless of their domain, as all of them exploit standard query features. This related work serves as the basis of the prompts that the proposed approach can provide to a tester when identifying MRs. 

\subsection{Automated Metamorphic Relation Generation for QBSs}
In 2022, Sergio et al.\cite{9826124} proposed an automated way of generating MRs through constraint programming\cite{9826124}; however, their proposed method requires the tester to provide a lightweight specification of the query parameters of the QBS, this process of specification inference might require extensive domain knowledge and may not be available or easy to induce. In addition, their proposed method also requires a source test case that contains no errors for their program to generate MRs. In comparison, the proposed method of this paper does not require such specification, nor does it need test cases with complete ground truth to work, arguably making it more applicable to real-world QBSs that require testing. 

\subsection{Enhancing Existing Testing Methods with MT}
In Sec. \ref{CS2}, we see the compatibility of the proposed approach and CT. The combination of MT and CT has been a rising topic of research, discussed in cases such as \cite{8411774},  as CT is often faced with the Oracle problem\cite{6963470}, which MT can alleviate\cite{6613484}. In this view, Enhance CT With Metamorphic Relations\cite{9629275} presents a CT methodology to enhance traditional CT by accounting for metamorphic relations; however, this method mainly focuses on the mix of CT and MT and assumes that the tester already has the MRs whereas the proposed approach prompts the tester to identify the MRs. 

%% file: 6-Conclusion.tex
\section{Conclusion} \label{ConclusionSec}

In this paper, an approach for prompting the identification of MRs for MT in real-world QBSs is proposed. The proposed approach is efficient in terms of manpower since the only process that requires extensive time and effort is the observation of the comparison table and identification of input relations. 

Furthermore, the proposed approach does not require ground truth for the QBS under test, alleviating the Oracle problem\cite{6963470} and making the application of this approach in the testing of real-world QBSs arguably more practical than methods that require source test cases or ground truth in order to test and validate a given QBS. 

Furthermore, through the conducted experiments, it can also be established that the proposed approach can be applied and successfully combined with other testing methods such as FT and CT, opening the door to more possibilities in terms of research on MT as well as improving the ease, efficiency, and effectiveness of QBS testing.